\documentclass[iop]{emulateapj}
\usepackage{graphicx}
\usepackage{amssymb,amsmath}
\usepackage{natbib}
\usepackage{epsfig}
\renewcommand{\vec}[1]{\mbox{\protect\boldmath$#1$}}

\newcommand{\Mc}{M_{\text{c}}}
\newcommand{\pf}{\rho_{\text{m}}}
\newcommand{\pe}{\rho_{\text{e}}}
\newcommand{\ephi}{{e_{\phi}}}

\newcommand{\vphi}{v_{\phi}}
\newcommand{\pmax}{P_{\text{max}}}

\newcommand{\psig}{\Psi_{\text{Sg}}}
\newcommand{\tpsig}{\tilde{\Psi}_{\text{Sg}}}
\renewcommand{\th}{\tilde{H}}
\newcommand{\tpsic}{\tilde{\Psi}_c}
\newcommand{\tphi}{\tilde{\Phi}}
\newcommand{\tm}{\tilde{\cal{M}}}

\newcommand{\aphi}{A_{\phi}}
\newcommand{\elik}{{K}}
\newcommand{\elie}{{E}}

\newcommand{\dt}{d_{\text{t}}}

\newcommand{\dift}{\partial_t}
\usepackage{float}

\begin{document}
\title{Electrically charged matter in permanent rotation around 
magnetized black hole\\
A toy model for self-gravitating fluid tori}

\author{A.Trova, V. Karas}
\affil{Astronomical Institute, Czech Academy of Sciences,\\ 
Bo\v{c}n\'{\i} II 1401, Prague, CZ-141\,31, Czech~Republic}

\and

\author{P. Slan\'{y}, J. Kov\'{a}\v{r}}
\affil{Institute of Physics, Faculty of Philosophy and Science, Silesian University in Opava\\
Bezru\v{c}ovo n\'{a}m. 13, CZ-746\,01 Opava, Czech Republic}




\begin{abstract}
We present an analytical approach for the equilibrium of a self-gravitating charged fluid embedded in a spherical gravitational and dipolar magnetic fields produced by a central mass. Our scheme is proposed, as a toy-model, in the context of gaseous/dusty tori surrounding supermassive black holes in galactic nuclei. While the central black hole dominates the gravitational field and it remains electrically neutral, the surrounding material has a non-negligible self-gravitational effect on the torus structure. By charging mechanisms it also acquires non-zero electric charge density, so the two influences need to be taken into account to achieve a self-consistent picture. With our approach we discuss the impact of self-gravity, represented by the term $\dt$ (ratio of the torus total mass to the mass of the central body), on the conditions for existence of the equilibrium and the morphology and typology of the tori. By comparison with a previous work without self-gravity, we show that the conditions can be different. Although the main aim of the present paper is to discuss a framework for the classification of electrically charged, magnetized, self-gravitating tori, we also mention potential astrophysical applications to vertically stratified fluid configurations.

\end{abstract}


\section{Introduction}
Studies of equilibrium of toroidal structures of a perfect fluid are important to understand the physics of accretion disks in active galactic nuclei -- AGN \citep{KoJaAbra78,AbrJaSi78,AbraCuSch84} and the dense self-gravitating tori around stellar mass black holes, which can be the result of the merger of a black hole -- neutron star binary or a remnant after the collapse of a massive star \citep{Shibata07,OtaTakaEri09,FuTaYosh13}. Nuclei of galaxies contain dusty tori and a central compact body that is frequently associated with a supermassive black hole (the mass typically $M_c\simeq10^6$--$10^9M_\odot$ \citep{Krolik14,EckSchStr05}). At a distance of $10^4$--$10^5$ gravitational radii ($R_g\equiv GM_c/c^2\approx1.5(M_c/M_\odot)\;$km, where $G$ is the gravitational constant) these tori become self-gravitating \citep{Cohu01,Karas04}. At the same time this distance is large enough to reduce the effects of General Relativity (essential near the center) to negligible level \citep{ShlosBegel87,hure98}. Therefore, an adequate and accurate description of the torus can be made using the fluid equations with Newtonian gravity \citep{FraKinRai85}. This subject has been studied in great detail \citep{StuSlaHle00,FoDai02,KucSlStu11,slany13,KovarTr14} also within the framework of General Relativity.

On the other hand, at smaller distances from the central body, the role of self-gravity of the fluid becomes less important relative to other forces. Here we will discuss a transitional region; we will take self-gravity into account while maintaining the Newtonian description of the central field to which we add as a new ingredient the effect of large-scale magnetic dipole field attached to the central body. The role of self-gravity is commonly described by Toomre's criterion \citep{Toomre64,goly65}. The self-gravity of toroids plays an important role because it affects the shape of the equilibrium configurations. Figures of rotating self-gravitating fluids have been described in great detail since decades under various assumptions. \cite{OstMa68,clement74,EriMu85} and \cite{hachisu86} studied self-gravitating systems in rotation without magnetic field. \cite{Eriguchi05} and \cite{YoEri06} included the effect of poloidal and toroidal magnetic field in rotating magnetized stars. Studies of stability have been performed too \citep{MaNiEri98,LuChenYaZa00}. This work has been extented to magnetized torus-central compact object systems by \cite{OtaTakaEri09} and \cite{FuTaYosh13}. The morphology of the solutions, their equilibrium and stability depend on many factors such as the rotation law, or the polytropic index. Many sequences have been found as for instance the Maclaurin, the Jacobi or the one-ring \citep{hachisu86,ansorg03,Eriguchi05}. These configurations of rotating fluids have also been studied within general relativity framework, with both analytical and numerical approaches \citep{Lanza92,NiEri94,LuChenYaZa00,Shibata07,RezBaioGia10}.

The torus acts as a source of material that gradually sinks from the outer regions down to the core, where it becomes heated and then accreted, typically on the viscous time-scale. This mechanism drives accretion and it helps to transfer the angular momentum and releases energy in the form of outflowing winds and radiation. High-energy X-rays originate near the inner rim of the accretion flow. The temperature of X-ray illuminated dust particles grows until they sublimate \citep{CzeHry11}. During the photo-ionization the dust grains develop electric charges attached to their surface, and so a complex dusty plasma is formed \citep{Horanyi96,VlaTsyMor05}. 

A central body is necessary to maintain the gravitational stability of an accretion torus. Black holes in centres of galaxies are generally believed to be electrically neutral because of selective charge accretion of the ambient plasma \citep{Kovar08,Kovar11}. A small non-zero charge is possible when external magnetic fields interact with a rotating body \citep{Wald74}. However, the surrounding dust particles may indeed keep non-vanishing charges while the system as a whole remains in neutral equilibrium \citep{DraSal79,Horanyi96,Swamy05}. Let us note that combined influences of magnetic fields and self-gravity should be included not only for the sake of completeness and consistency of the model. Their impact may be essential for the structure of accretion flows and the final rate of mass accretion. It has been argued that magnetized accretion disks are less susceptible to gravitational fragmentation \citep{SalSiArmBeg16,PaBlaBol03}.Recently, in a different context of geometrically thin accretion disks, \cite{Scadowski16} examined the role that strong large-scale magnetic fields threading the fluid can play on thermal stabilization of the flow.

In this paper, we apply the Newtonian hydrodynamical approach. The torus is modeled by a perfect fluid with some net electric charge spread through the fluid. This model represents a different limit to the well-known ideal magnetohydrodynamics (MHD) with zero resistivity and vanishing volume charge density. The  approximation  of  ideal MHD is  accurate  in  many  astrophysically-relevant situations involving  fluids in motion \citep{Melrose80}. In contrast to this approach, we are working with zero-conductivity, meaning a non vanishing electric charge density of the fluid distribution, such as ionized plasma \citep{WaNg99,PanWar08,InouInu08}. Despite the fact that the net charge is negligible in majority of astrophysical realistic systems, where the accreted material is described in terms of ideal fluid with high conductivity that satisfies the conditions of force-free limit, there has been an ongoing debate about the role of charge separation that may be caused by processes operating in complex (two-component) cosmic plasmas. These include the influence of irradiation by photons emanating from the central source, or the effect of electric forces acting parallel to the magnetic field lines in the local co-moving frame of the magnetized plasma, see \ref{sec:Nonzero} for more discussions.   

We present a convenient formalism and we give examples of a viable physical set-up where both the self-gravity of the torus material and non-vanishing electric charge density interact to define the radial and vertical structure of an equilibrium configuration torus. The idea is to use the same method as \cite{slany13}. They work with four families of specific charged distributions and a general form for the orbital velocity. In addition to \cite{slany13}, we include effects of self-gravity, which has been previously neglected for simplicity. Here we decided to work, for convenience, with two of these four families, namely family II and IV. These two distributions have the advantage to produce equatorial and off-equatorial solutions, so they capture the two qualitatively different cases.

The paper is organized as follows. In section \ref{sec:section1}, we present the basic equations, the assumptions made, the normalization and the general conditions of equilibrium existence. Section \ref{sec:equ_tori} is dedicated to the study of equatorial tori. We show how the self-gravity influences the conditions of existence of the tori, their charge and their morphology. The same study is done for off-equatorial tori in section \ref{sec:off_cond}. In section \ref{sec:SCF} we employ the numerical method of Self-Consistent Field to compare the precision of our analytical approximation with a corresponding solution, where some of the restricting assumptions can be relaxed. Conclusions are given in section \ref{sec:conclusion}.

\section{Basic equations and hypotheses}
\label{sec:section1}
\label{sec:basic}
The tori equilibrium condition is governed by the Euler equation (\cite{LanLif87}, equation ($2.3$)) in its stationary form, in which we have added two terms - the first one describing the self-gravitation and the second one corresponding to the Lorentz force density (and describing the electromagnetic interaction of charged fluid with external (electro)magnetic field. In the Newtonian limit, the equation adopts the following form:
\begin{flalign}
\label{eq:Euler}
\pf(\dift v_{i}+v^{j}\nabla_j v_{i})=&-\nabla_{i}P- \pf \nabla_{i}\Psi - \pf \nabla_{i}\psig \\ \nonumber
&+ \rho_{e}(E_{i}+\epsilon_{ijk}v^{j}B^{k}), 
\end{flalign}
where $P$, $\Psi_{Sg}$, $\Psi_c$, $v$, $\pf$ and $\pe$ are, respectively, pressure, the self-gravitational potential of the torus, the central mass potential, velocity of the fluid, the mass density and the charge density. All quantities are functions of cylindrical coordinates $R$ and $Z$. The electromagnetic field is described by its electric part E and magnetic part B. The last term on the right hand side of equation (\ref{eq:Euler}) corresponds to the Lorentz force.

The charged gas velocity is assumed to be the same as the fluid velocity, so the conservation of mass and electric charge are described by the following continuity equations:
\begin{subequations}
\label{eq:continuity}
\begin{align}
&\dift \pf +\nabla_i (\pf v^{i}) = 0,\\
&\dift \pe +\nabla_i (\pe v^{i}) =0.
\end{align}
\end{subequations}
In our work, the fluid is assumed to be stationary, axially symmetrical, self-gravitating, and embedded in a spherical gravitational and dipolar magnetic field, so
\begin{flalign}
E_i=0, \quad i=(R,\phi ,Z), \quad \dift=0.
\end{flalign}
Due to our opposite approach to the ideal MHD (i.e. assuming vanishing  conductivity), it is reasonable to prescribe the azimuthal motion of the fluid only (and no meridional or radial one), so
\begin{flalign}
v_R=v_Z=0, \quad \vphi=\vphi(R,Z).
\end{flalign}
Thus, equations (\ref{eq:continuity}a) and (\ref{eq:continuity}b) are fulfilled automatically. Equation (\ref{eq:Euler}) can be rewritten as:
\begin{equation}
\label{eq:bern}
-\frac{1}{\pf} \vec{\nabla}{P}-\vec{\nabla}{\Psi_{Sg}}- \vec{\nabla}{\Psi_c}-\vec{\nabla}{ \Phi}+\frac{\vec{\cal{L}}}{\pf}=0,
\end{equation}
where $\Phi$ is the centrifugal potential and $\vec{\cal{L}}$ the Lorentz force. Equation (\ref{eq:bern}) holds for an isentropic case of ideal fluid. It can be integrated and rewritten as:
\begin{equation}
H+\Psi_{Sg}+\Psi_c+\Phi+\cal{M}=\text{Const},
\label{eq:bernouilli}
\end{equation}
where $H$ is the enthalpy and $\cal{M}$ the ``magnetic potential". Those quantities can be expressed, using cylindrical coordinates  ($R$,$\phi$,$Z$), as 
\begin{equation}
      \begin{aligned}
        H&=\int \frac{dP}{\pf}, \quad
        \Psi_c=\frac{-G\Mc}{\sqrt{R^2+Z^2}},\\
        \Phi&=-\int \frac{v_{\phi}^2}{R} dR, \quad
        -\vec{\nabla}{\cal{M}}=\frac{\vec{\cal{L}}}{\pf},\\
      \end{aligned}
\end{equation}
where $M_c$ is the mass of the central object. The self-gravitational torus potential is given by Poisson's equation:
\begin{equation}
\Delta \psig = 4\pi G \pf.
\end{equation}
In our work $\psig$ is approximated by the gravitational potential of a loop  in the equatorial plane, coordinates ($r_c$,$0$) and mass $m$, centred on the axis. In cylindrical coordinates ($R$,$\phi$,$Z$), it is given by \citep{durand64}:
\begin{equation}
\Psi_{Sg} \sim -\frac{Gm}{r_c\pi}\sqrt{\frac{r_c}{R}}k\elik(k),
\end{equation}
with
\begin{equation}
k=\frac{2\sqrt{r_cR}}{\sqrt{(r_c+R)^2+Z^2}}.
\end{equation}
The complete elliptic integral of the first kind $\elik$ \citep{gradryz65} diverges when its modulus $k=1$ (i.e when the field point $(R,Z)$ coincides with the loop radius $(r_c,0)$). To avoid this singularity we add a parameter $\lambda$ to the modulus $k$. 
\begin{equation}
\label{eq:newk}
\frac{2\sqrt{r_cR}}{\sqrt{(r_c+R)^2+Z^2}} \rightarrow \frac{2\sqrt{r_cR}}{\sqrt{(r_c+R)^2+Z^2+\lambda^2}}
\end{equation}
This technique, initially developed to handle numerical N-body simulations, is used to compute the gravitationnal potential of gazeous self-gravitating disks \citep{paplin89,tremaine01,lietal08}. The free-parameter, $\lambda$, called ``the smoothing length", takes into account the vertical and the radial extension of the torus. Various prescriptions have been chosen for this parameter: (i) a function of the disk parameter,(ii) a function of space or (ii) a constant value \citep{masset02,hurepierens05,kley12,hutr15}, however no universal value has been adopted; see \cite{hp09} for an non exhaustive list. In our work, the softening length is a function of the loop radius (i.e location of the maximum pressure), $\lambda=0.4r_c$. Even if this parameter influences the value of the gravitational potential, the morphology of the solutions remains preserved.

The last potential of the four unknown potential functions to be determined in equation (\ref{eq:bernouilli}) is the ``magnetic potential" $\cal{M}$. As we know the Lorentz force is given by
\begin{equation}
\vec{\cal{L}}=\vec{j}\wedge\vec{B},
\end{equation} 
where $\vec{j}=\pe \vphi \vec{\ephi}$ and $\vec{B}$ are, respectively, the fluid's current density and the external magnetic field. The $\cal{M}$-function has to satisfy
\begin{equation}
\vec{\nabla}{\cal{M}}=-\frac{\vec{\cal{L}}}{\pf}.
\label{eq:satisfyM}
\end{equation} 
To solve this equation we impose proportionality 
\begin{equation}
\pe=\pf q(R,Z).
\end{equation} 
$\cal{M}$ depends on the rotation law and the specific charge distribution $q(R,Z)$, see Appendix \ref{appendixA} for explicit equations. $\cal{M}$ is the solution of the equation (\ref{eq:condition}). To solve the equilibrium equation, we set various hypotheses on the system.

\subsection{Assumptions}
\label{sec:asumptions}
We assume that:
\begin{enumerate}
\item The fluid is axially symmetrical and also symmetric with respect to the mid-plane.
\item  The fluid is incompressible, $\pf=\text{Const}$, so the enthalpy is then written as
\begin{equation}
\label{eq:polytropic}
H=\frac{P}{\pf}.
\end{equation}
\item The integrability condition of the equation (\ref{eq:bern}) leads to two unknown functions: the orbital velocity $v_{\phi}(R,Z)$, i.e. the way of rotation of the fluid, and the specific charge $q(R,Z)$.
\item The fluid is embedded in an external dipolar magnetic field, which is given in cylindrical coordinates by
\begin{equation}
\label{eq:b}
\left\{
\begin{aligned}
&B_R= \frac{3 \mu ZR}{\left(R^2+Z^2\right)^{5/2}},\\
&B_{\phi}=0,\\
&B_Z=\frac{\mu(2Z^2-R^2)}{\left(R^2+Z^2\right)^{5/2}}.
\end{aligned}
\right.
\end{equation}
\end{enumerate}
Relations (\ref{eq:b}) imply the following expression for the electromagnetic potential, $\aphi$,
\begin{equation}
\aphi=\frac{\mu R}{\left(R^2+Z^2\right)^{3/2}}.
\end{equation}
According to \cite{slany13}, analysis of integrability condition shows that $v_{\phi}=v_{\phi}(R)=K_2R^{K1}$ (where $K_1$ and $K_2$ are constants). So the centrifugal potential is
\begin{equation}
\Phi=-\frac{K_2^2}{2K_1}R^{2K_1}.
\end{equation}
The choice of specific charge distribution reflects the choice of the rotation law, in order to have an integrable system. We decided to study two of the four specific charge distributions described in \cite{slany13}, the family II and IV. We have in cylindrical coordinates $(R,\theta,Z)$ \\
\begin{equation}  
\label{eq:familyII}  
\left\{
\begin{aligned}  
        &q(R,Z)=C\frac{(R^2+Z^2)^{(3/4+3K_1/2)}}{R^{3K_1}},\\
        &{\cal{M}}(R,Z)=2\mu K_2C\frac{ \left(R^2+Z^2\right)^{3K_1/2-3/4}}{(2K_1-1)R^{2K_1-1}},
\end{aligned}
\right.
\end{equation}
for family II, and
\begin{equation}
\label{eq:familyIV}
\left\{
\begin{aligned}
        &q(R,Z)=C\left(\frac{R^2}{R^2+Z^2}\right)^{3(1-K_1)/2},\\
        &{\cal{M}}(R,Z)=\mu K_2C\frac{R^{4-2K_1}}{(K_1-2)\left(R^2+Z^2\right)^{3-3K_1/2}},
        \end{aligned}
\right.
\end{equation}
for family IV.
In the following examples, we set the model parameters in such a way that the imposed central dipole field dominates over the magnetic field produced by the current of the charged rotating tori. This assumption is checked in the section \ref{sec:equa_conf}.

\subsection{Normalization}
We introduce dimensionless physical quantities denoted by ``tilde''. For the normalization, we use various quantities: $X=R/r_c$, $Y=Z/r_c$, $\pf$, $\pmax$, $r_c$, $G$, $C$, $\mu$ and $K_2$. The dimensional variables are given by

\begin{equation}
\begin{aligned}
        &\psig=\tpsig Gm/r_c, \\
        &\Psi_c=\tpsic GMc/r_c,\\
        &\phi=\tphi K_2^2r_c^{2K_1},\\
        &H= \pmax\th/\pf=a\tilde{H},\\
       &\cal{M}= \left\{
 \begin{aligned}
       & \tm \mu C K_2r_c^{K_1-1/2} \quad \text{(II)},\\ 		   		        &\tm \mu C K_2r_c^{K_1-2}\quad \text{(IV)}.
   \end{aligned}
    \right.
    \end{aligned}
\end{equation}
With these new variables, the equation (\ref{eq:bernouilli}) becomes
\begin{equation}
a\th + \dt \tpsig + \tpsic + b\tphi +e\tm=c,
\label{eq:bern_norm}
\end{equation}
with
\begin{equation}
  \left\{
      \begin{aligned}   
        &a=\frac{\pmax r_c}{\pf G\Mc},  \\ 
        &b=\frac{K_2^2 r_c^{2K_1+1}}{G\Mc}, \\
        &c=\frac{\text{Const}r_c}{G\Mc},\\
        &\dt=\frac{m}{\Mc},\\
        \end{aligned}
        \right.
        \quad
        e=\left\{
        \begin{aligned}
        &\frac{\mu K_2 Cr_c^{K_1+1/2}}{G\Mc}\quad \text{(II)},   \\ 		&\frac{\mu K_2 Cr_c^{K_1-1}}{G\Mc} \quad \text{(IV)}.\\
        \end{aligned}
    \right.
\end{equation}
The $c$-constant determines the surface of zero pressure (i.e the boundary of the torus). 

\subsection{General conditions for the existence of equilibrium.}
An equilibrium solution exists if there is a local pressure maximum, or a local enthalpy maximum. We suppose that the maximum is located in $(R=r_c,Z=z_c)$, i.e in $(X=1,Y=Y_c)$. The necessary condition to have a local extremum in this point is 
\begin{equation}
\label{eq:general_cond}
\nabla \th=0.
\end{equation}
This extremum corresponds to a maximum if 
\begin{equation}
 \left\{
 	\begin{aligned}
&\frac{\partial^2 \th}{\partial X^2}<0 \quad \text{and}\\
&\frac{\partial^2 \th}{\partial X^2} \times \frac{\partial^2 \th}{\partial Y^2}- \left(\frac{\partial \th}{\partial X \partial Y}\right)^2 >0.
	\end{aligned}
 \right.
\label{eq:max_conditions}
\end{equation}
The expressions of these derivatives are given in the Appendix \ref{appendixC}. 

In the following sections, we will show that there are both equatorial toroidal configurations, where the maximum pressure takes place in $(X=1,Y=0)$ and off-equatorial toroidal configurations, where the maximum pressure is placed in $(X=1,Y=Y_c)$.

\section{Equatorial tori: Incompressible fluid}
\label{sec:equ_tori}
For equatorial tori, the maximum, located at ($X$,$Y$)=($1,0$) must satisfy equations (\ref{eq:general_cond}) and (\ref{eq:max_conditions}). For both distributions of specific charge, the maxima of pressure exists if $b$, $\dt$ and $e$ satisfy the following conditions.
\begin{itemize}
\item Family II
\begin{subequations}\\
 \label{eq:cond_familyII}
      \begin{align} 
        &\text{If} \quad K_1+\frac{1}{2}\gtrless 0 \quad \text{then} \label{eq:existence1_distrib1} \\ \nonumber
        &b \lessgtr -1+ \frac{\dt}{K_1+1/2}\left[ \frac{\partial^2 \tpsig}{\partial X^2} -\frac{2K_1-3}{2} \frac{\partial \tpsig}{\partial X}\right]\\ 
        &\text{and}\\
        &b > \frac{2}{3} - \frac{\dt}{3}\left[ \frac{\partial^2 \tpsig}{\partial Y^2} -3 \frac{\partial \tpsig}{\partial Y}\right]\label{eq:existence2_distrib1},\\
        &e=b-\dt\frac{\partial \tpsig}{\partial X}-1.
      \end{align}
\end{subequations}

\item Family IV
\begin{subequations}
\label{eq:cond_familyIV}      
      \begin{align}
        &\text{If} \quad (K_1+2)\gtrless 0 \quad \text{then} \label{eq:existence1_distrib2} \\ \nonumber
        &b \lessgtr \frac{1-K_1}{K_1+2}+ \frac{\dt}{K_1+2}\left[ \frac{\partial^2 \tpsig}{\partial X^2} -(K_1-3) \frac{\partial \tpsig}{\partial X}\right]\\\nonumber\\ 
         &\text{and}\\
        &b > \frac{2}{3} - \frac{\dt}{3}\left[ \frac{\partial^2 \tpsig}{\partial Y^2} -3 \frac{\partial \tpsig}{\partial Y}\right],\label{eq:existence1_distrib2c}\\
        \nonumber\\
        &e=b-\dt\frac{\partial \tpsig}{\partial X}-1,
      \end{align}
\end{subequations}

\end{itemize}
where the value of $\partial \tpsig/\partial X$, $\partial^2 \tpsig/\partial X^2$ and $\partial^2 \tpsig/\partial Y^2$ are calculated in $(X,Y)=(1,0)$. Their analytical expressions are given in Appendix \ref{appendixB}. The conditions described above depend on the mass ratio between the central mass and the torus mass, on the rotation law (i.e. the specific charge) and on the gradient and laplacian of the self-gravitational field of the torus. As shown in the section \ref{sec:basic}, the latter is approximated by the gravitational potential of the ring, located in ($1,0$) with a good accuracy. So the conditions given above are valid to the first order. They can give us information on the role of the self-gravity, on the conditions of equilibrium and on the configuration of tori. We can play with various parameters $\dt,e,b$ and $r_c$.

\subsection{Influence of self-gravity on the equilibrium conditions}
The conditions vary with the rotation law, so they are functions of $K_1$. For definiteness, here we are going to study two different rotation laws: a rotation with constant angular momentum, $K_1=-1$, and the rigid rotation (i.e constant angular velocity), $K_1=1$.

\subsubsection{Constant angular momentum: $K_1=-1$.}
\label{sec:conditions_angular}
\begin{figure*}
\centering
  \includegraphics[width=1\hsize]{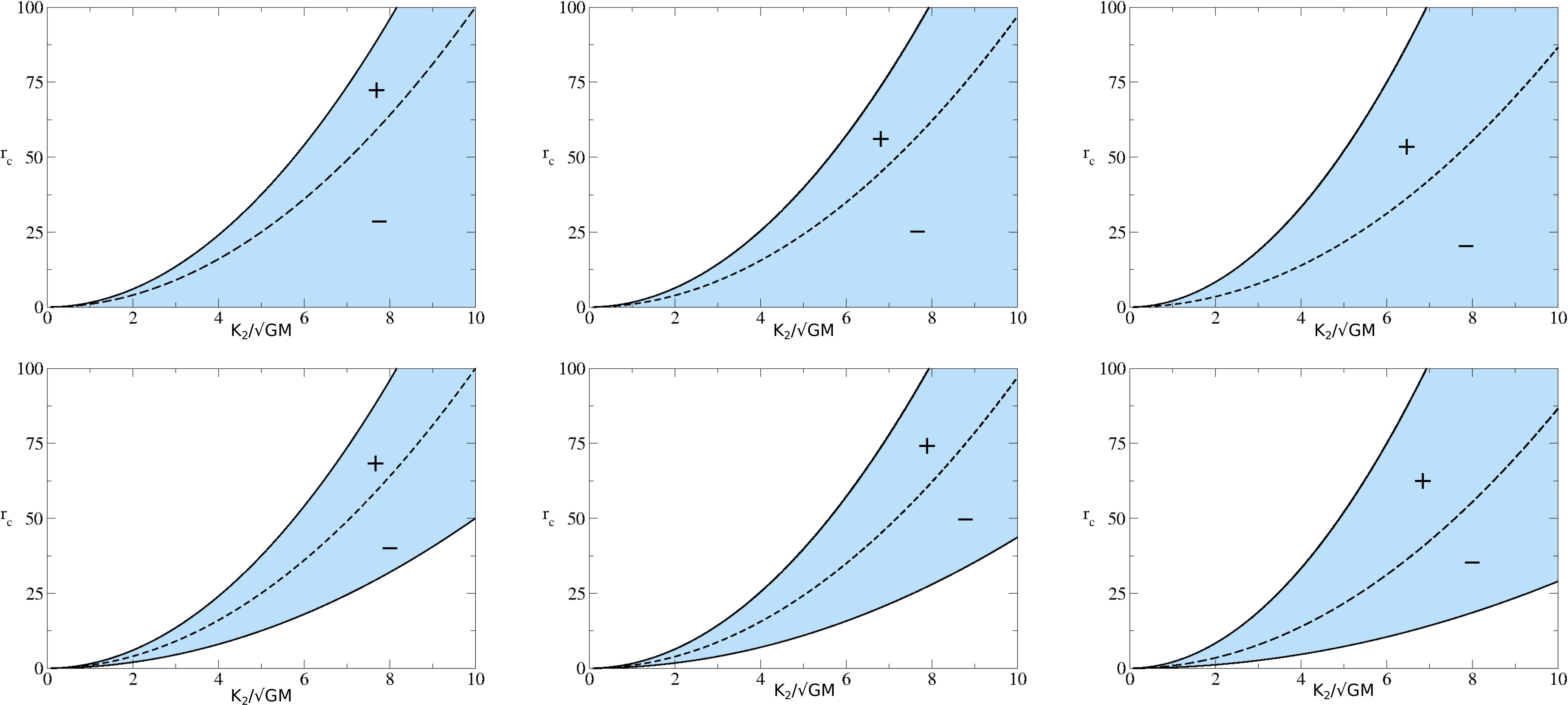}
	\caption{Representation of conditions of existence: panels at the top are for family II, and panels at the bottom for family IV. From the left to the right, the value of $\dt$ raises, $\dt=0,0.1,0.5$. This work is done for case of rotation with constant angular momentum, $K_1=-1$.}
  \label{f1}
\end{figure*}

In the case where the self-gravity is neglected, $\dt=0$, for the family II and IV, the conditions (\ref{eq:cond_familyII}) and (\ref{eq:cond_familyIV}) are, respectively, graphically represented in Figure \ref{f1}. Now, we add the self-gravity by giving a positive value to $\dt$. The result is shown in the Figure \ref{f1} at the top for family II and at the bottom for family IV. We chose two different values for $\dt$: $\dt=0.1$ (in the middle), the torus mass represents one percent of the central mass and $\dt=0.5$ (on the right). We can clearly see the influence of the self-gravity on the conditions. For both distributions, the range rises. If we impose $K_2/\sqrt{GM}=2.5$, the range of possible value of $r_c$ changes. It is shown in the Table \ref{tab1} for both families.
\begin{table}
\begin{center}
\caption{Range of possible values of $r_c$ for three values of $\dt$.\label{tab1}}
\begin{tabular}{lll}
\hline
\hline
\multicolumn{3}{c}{Family II}\\
\hline
 $\dt$ & $K_2/\sqrt{GM}$ & \text{range of} $r_c$ \\
\hline
   $0$& $2.5$ & $[0,9.37]$ \\
   \hline
   $0.1$ & $2.5$ & $[0,9.93]$\\
   \hline
   $0.5$ & $2.5$ & $[0,13]$\\
   \hline
\end{tabular}
\begin{tabular}{lll}
\hline
\hline
\multicolumn{3}{c}{Family IV}\\
\hline
 $\dt$ & $K_2/\sqrt{GM}$ & \text{range of} $r_c$ \\
\hline
   $0$& $2.5$ & $[3.12,9.37]$ \\
   \hline
   $0.1$ & $2.5$ & $[2.73,9.93]$\\
   \hline
   $0.5$ & $2.5$ & $[1.81,13]$\\
   \hline
\end{tabular}
\end{center}
\end{table}

\subsubsection{Rigid rotation: $K_1=1$.}
\begin{figure}
\centering
  \includegraphics[width=1.\hsize]{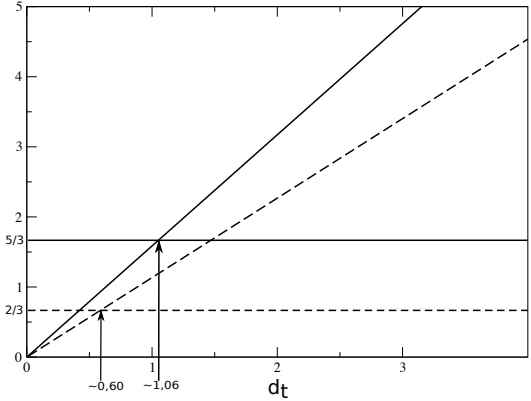}
	\caption{Plot of right hand sides of inequality (\ref{eq:cond_rigid_II}) (full line) and inequality (\ref{eq:cond_rigid_IV}) (dashed line).}
  \label{f2}
\end{figure}

For the case of a rigid rotation, $K_1=1$, if we neglect the self-gravity, i.e. $\dt=0$, the conditions (\ref{eq:cond_familyII}) and (\ref{eq:cond_familyIV}) are not satisfied, as seen in \cite{slany13}. No equilibrium configuration of equatorial tori can be found for this specific rotation law. In the case where $\dt \neq 0$, the conditions (\ref{eq:cond_familyII}b) and (\ref{eq:cond_familyIV}b) become
\begin{itemize}
\item Family II
\begin{subequations}\\
 \label{eq:cond_familyII_rigid}
      \begin{align} 
        &b < -1+ \frac{2\dt}{3}\left[ \frac{\partial^2 \tpsig}{\partial X^2} +\frac{1}{2} \frac{\partial \tpsig}{\partial X}\right],\\ 
        &b > \frac{2}{3} - \frac{\dt}{3}\left[ \frac{\partial^2 \tpsig}{\partial Y^2} -3 \frac{\partial \tpsig}{\partial Y}\right].\label{eq:existence1_distrib1_rigid}
      \end{align}
\end{subequations}
\item Family IV
\begin{subequations}
\label{eq:cond_familyIV_rigid}      
      \begin{align}
        &b < \frac{\dt}{3}\left[ \frac{\partial^2 \tpsig}{\partial X^2} +2 \frac{\partial \tpsig}{\partial X}\right],\\\nonumber\\ 
        &b > \frac{2}{3} - \frac{\dt}{3}\left[ \frac{\partial^2 \tpsig}{\partial Y^2} -3 \frac{\partial \tpsig}{\partial Y}\right].\label{eq:existence2_distrib1_rigid}
      \end{align}
\end{subequations}

\end{itemize}
For family II, the conditions above are valid if 
\begin{equation}
\frac{5}{3}<\frac{\dt}{3}\left(\frac{\partial^2 \tpsig}{\partial Y^2}+2\frac{\partial^2 \tpsig}{\partial X^2}-2\frac{\partial \tpsig}{\partial X}\right),
\label{eq:cond_rigid_II}
\end{equation}
and for family IV, if
\begin{equation}
\frac{2}{3}<\frac{\dt}{3}\left(\frac{\partial^2 \tpsig}{\partial Y^2}+\frac{\partial^2 \tpsig}{\partial X^2}-\frac{\partial \tpsig}{\partial X}\right),
\label{eq:cond_rigid_IV}
\end{equation}
where the value of $\partial \tpsig/\partial X$, $\partial^2 \tpsig/\partial X^2$ and $\partial^2 \tpsig/\partial Y^2$ are calculated in $(X,Y)=(1,0)$. The right hand side of inequality (\ref{eq:cond_rigid_II}) (full line) and inequality (\ref{eq:cond_rigid_IV}) (dashed line) are plotted as a function of $\dt$ and compared to the left hand side in the Figure \ref{f2}.

We see that the conditions (\ref{eq:cond_rigid_II}) and (\ref{eq:cond_rigid_IV}) are valid when the parameter $\dt$ is greater than $1.06$ and $0.6$, respectively. It means that solutions in rigid rotation can exist, for these families, if the strength of the self-gravity is high enough. It is an interesting result because in the case of self-gravitating tori without a magnetic field and central mass, such solutions in rigid rotation exist too \citep{hachisu86}. On the other hand, also in the case of test fluid orbiting in central gravitational and dipolar magnetic fields, it is possible to find specific charge distributions (other than families II and IV studied there) allowing stationary equilibrium torus in rigid rotation.

\subsection{Influence of self-gravity on the tori equilibrium of both distributions}
\label{sec:equa_conf}
\begin{figure*}
\centering
  \includegraphics[width=1\hsize]{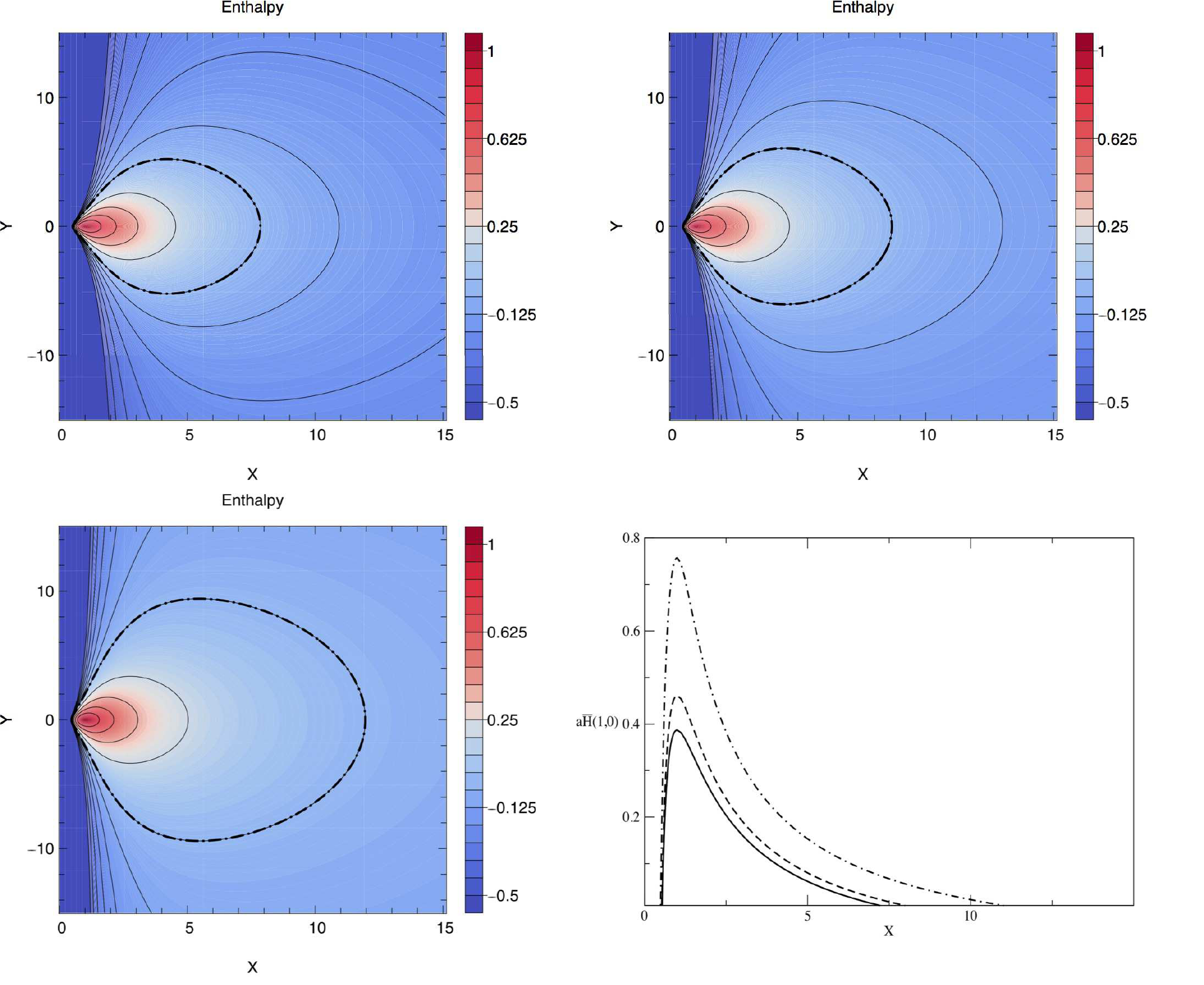}
	\caption{Maps of enthalpy distribution in positively charged tori ($\th$) for  $\dt=0$ (top left), $\dt=0.1$ (top right) and in negatively charged tori for $\dt=0.5$ (bottom left) for Family II. The parameter used to plot these graphs are given in Table \ref{tab2}. At the bottom right, the corresponding equatorial pressure profiles are shown (full line for $\dt=0$, dashed line for $0.1$ and dash-dot line for $0.5$).}
  \label{f3}
\end{figure*}

To show the influence of self-gravity, we choose three configurations with the same rotation law, specific charge distribution and location of maximum of pressure, and we vary the value of $\dt=[0,0.1,0.5]$. For the family II, according to the section \ref{sec:conditions_angular}, if we choose the pair $(r_c=6,K_2/\sqrt{GM}=2.5)$ then equilibrium is possible. We perform three tests, described in Table \ref{tab2}. They are graphically presented in Figure \ref{f3}. We choose for all of them $c=-0.02r_c$. This constant $c$ determines the surface where the pressure is equal to zero. It defines the boundary and the shape of the torus. For family II, we can see that the morphology of the solution does not change. The pressure field has a toroidal shape for all the three figures while the specific charge has a cylindrical topology. The main change appears in the charge of the torus. For $\dt=0$ and $\dt=0.1$ the torus is positively charged but it is negatively charged for $\dt=0.5$. This information is given by the value of $\mu C/\sqrt{GM}$ in Table \ref{tab2}. Another interesting effect is that the maximum of pressure raises with the value of $\dt$. The torus grows with the strength of the self-gravity. To check our assumption about the dominance of the dipolar field over the magnetic field produced by the current, we compare their magnitude close to the outer edge (closer to the center the central dipole component grows stronger). For the torus configuration given by the test in Table \ref{tab2} and \ref{tab3}, we compute by numerical integration the total electric charge. We estimate the angular velocity at the pressure maximum and we calculate the magnetic field of a rotating narrow charged ring with the same charge and the same angular velocity. We obtain, for all the tests prensented in the paper, $B_{\text{torus}}/B_{\text{dipole}} = 10^{-1} \sim 10^{-2}$. The dipole field is stronger than the one produced by the torus.

\begin{table}
\begin{center}
\caption{Three tests for family II with $K_2/\sqrt{GM}=2.5$ and $r_c=6$.\label{tab2}}
\begin{tabular}{ll}
\hline
\hline
   Test 1 \quad  $\dt=0$&  $\mu C/\sqrt{GM} \sim 0.041$ \\
   \hline
   Test 2 \quad $\dt=0.1$ & $\mu C/\sqrt{GM}\sim 0.010$\\
   \hline
   Test 3 \quad $\dt=0.5$ & $\mu C/\sqrt{GM}\sim -0.113$\\
   \hline
\end{tabular}
\end{center}
\end{table}

\begin{figure*}
\centering
  \includegraphics[width=1\hsize]{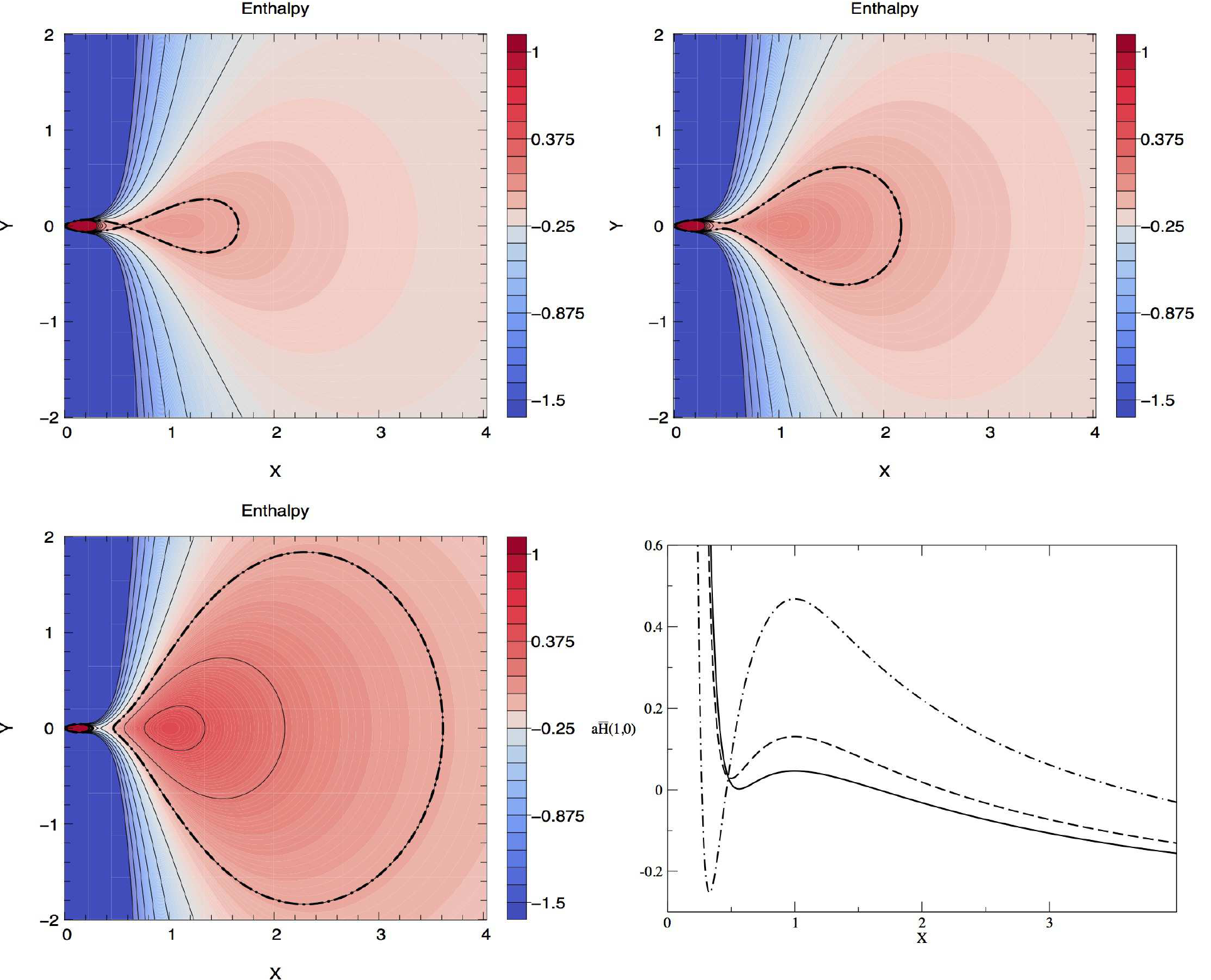}
	\caption{The same as in Figure \ref{f3}, but for family IV and the parameters from Table \ref{tab3}. However, two differences are included: 1) For each value of $\dt$, the torus is positively charged. 2) the morphology changed with the value of $\dt$ with the appearance of a cusp in the figure at the bottom left.}
  \label{f4}
\end{figure*}

For family IV, we perform the same test but with $r_c=4$, $K_2/\sqrt{GM}=2.5$ and $c=-0.09r_c$, see Table \ref{tab3} for the value of $\mu C/\sqrt{GM}$ for each test. We can see that the topology has changed. There is the formation of a cusp which corresponds to the minimum of the pressure. Resulting configurations are represented in Figure \ref{f4}. The effect of the self-gravity on the increase of central pressure and the extension of the torus seems to be the same as in the case of family II.
\begin{table}
\begin{center}
\caption{Three tests for family IV with $K_2/\sqrt{GM}=2.5$ and $r_c=4$.\label{tab3}}
\begin{tabular}{ll}
\hline
\hline
   Test 1 \quad $\dt=0$&  $\mu C/\sqrt{GM} \sim 3.6$ \\
   \hline
   Test 2 \quad $\dt=0.1$ & $\mu C/\sqrt{GM}\sim 3.4$\\
   \hline
   Test 3 \quad $\dt=0.5$ &  $\mu C/\sqrt{GM}\sim 2.6$\\
   \hline
\end{tabular}
\end{center}
\end{table}

\section{Off-equatorial tori: incompressible case}
\label{sec:off_cond}
\begin{figure*}
\centering
  \includegraphics[width=1\hsize]{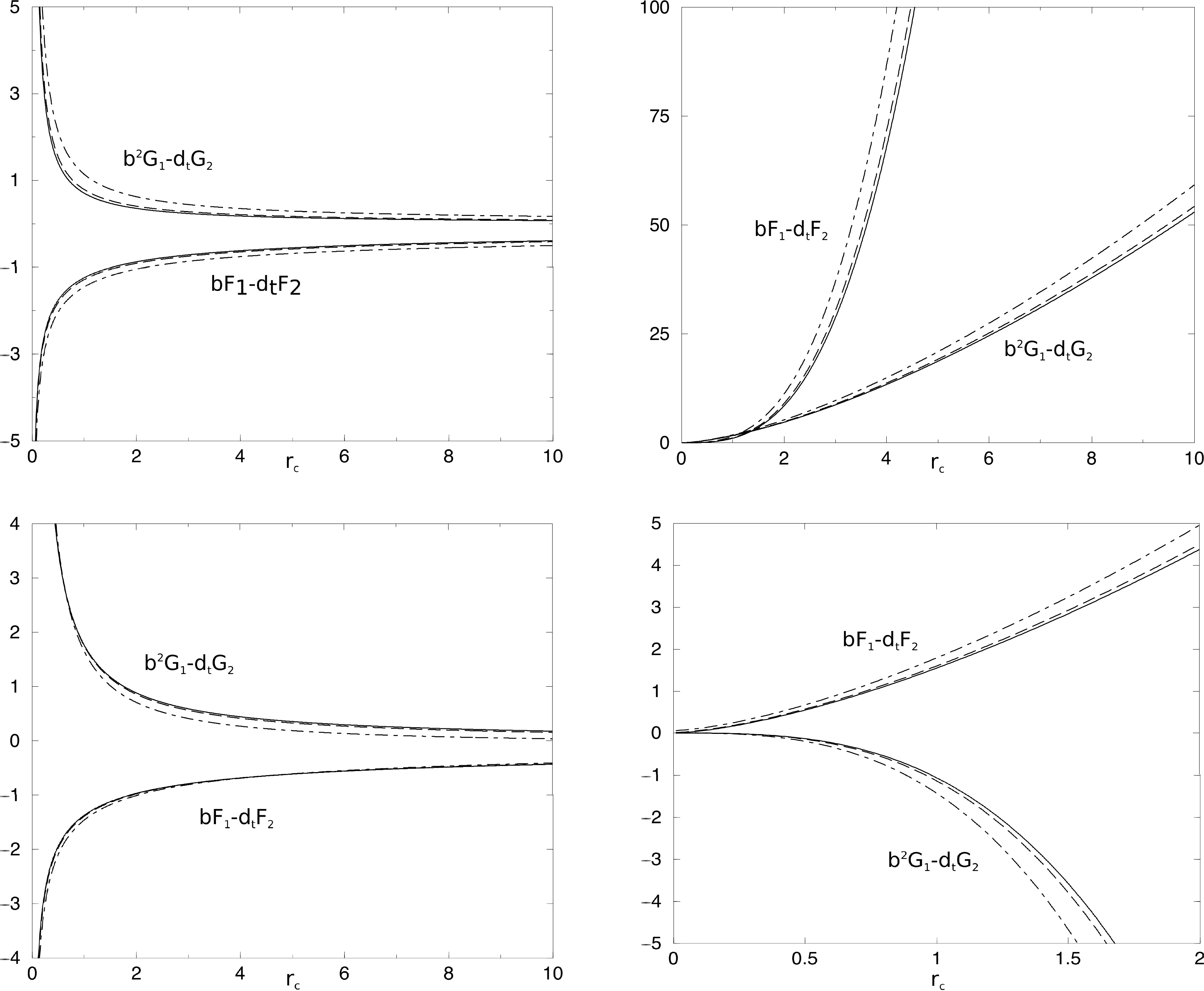}
	\caption{Representation of left hand-sides of equations (\ref{eq:off_condition1}) and (\ref{eq:off_condition2}) as a function of $r_c$. Tori of family II are represented at the top, with the case $K_1=-1$ at the left and $K_1=1$ at the right. Family IV is shown at the bottom.}
  \label{f5}
\end{figure*}

We perform the same study as previously. We suppose now that the maximum is located in $(1,Y_c)$, but we keep the ring in $(1,0)$ which simplifies the equations for the conditions of the torus existence. First, we search these conditions for both specific charge distributions (equation (\ref{eq:familyII}) and (\ref{eq:familyIV})). The value of the constant $b$ is not free, as before, but is now fixed by the conditions (\ref{eq:general_cond}) and (\ref{eq:max_conditions}) and is the same for both families:
\begin{equation}
b=\frac{2}{3\sqrt{1+Y_c^2}}+\dt\left[\frac{\partial \tpsig}{\partial X}-\frac{\partial \tpsig}{\partial Y}\frac{1-2Y_c^2}{3Y_c}\right].
\end{equation}
According to equation (\ref{eq:max_conditions}), the conditions are given by the following inequalities,
\begin{equation}
b F_1(1,Y_c)-\dt F_2(1,Y_c)<0,
\label{eq:off_condition1}
\end{equation}
and
\begin{equation}
b^2 G_1(1,Y_c)-\dt G_2(1,Y_c)>0.
\label{eq:off_condition2}
\end{equation}
The form is the same for both family II and IV. $F_1$, $F_2$, $G_1$ and $G_2$ are complicated functions of $Y_c$ and the first, second and mixed derivatives of the self-gravitational potential $\tpsig$. The expressions of these functions depends on the family, and are shown in Appendix \ref{appendixD}. The constant $e$ is given, for family II and IV respectively, by

\begin{subequations}
\label{eq:cond_familyIV_rigide}     
      \begin{align}
&    e=-\frac{\left[b-\dt\left(\frac{\partial \tpsig}{\partial X}-\frac{1}{Y_c}\frac{\partial \tpsig}{\partial Y}\right)\right]}{2\left(1+Y_c^2\right)^{3K_1/2-3/4}}, \\\nonumber\\ 
&        e=-\frac{\left[b-\dt\left(\frac{\partial \tpsig}{\partial X}-\frac{1}{Y_c}\frac{\partial \tpsig}{\partial Y}\right)\right]}{2\left(1+Y_c^2\right)^{3K_1/2-3}} 
        \label{eq:value_e}.
      \end{align}
\end{subequations}

and is always negative. Then for $K_2>0$ (positive rotation) and $\mu>0$ (given the orientation of the magnetic field), there are only negatively charged off-equatorial toroidal configurations, as in \cite{slany13}. Choosing loci of torus center at $Y_c=1$ for both families, we plot the conditions (\ref{eq:off_condition1}) and (\ref{eq:off_condition2})  for $K_1=-1$ and $K_1=1$ in the Figure \ref{f5}.

We can see that, as the equatorial case, for these two families, solutions with a rigid rotation law do not exist because the conditions (\ref{eq:off_condition1}) and (\ref{eq:off_condition2}) are not satisfied (see Figure \ref{f5} at the right). The main difference with the equatorial case is that there is always a solution for $K_1=-1$. The conditions seem to be satisfied everywhere. 

\begin{figure*}
\centering
  \includegraphics[width=1\hsize]{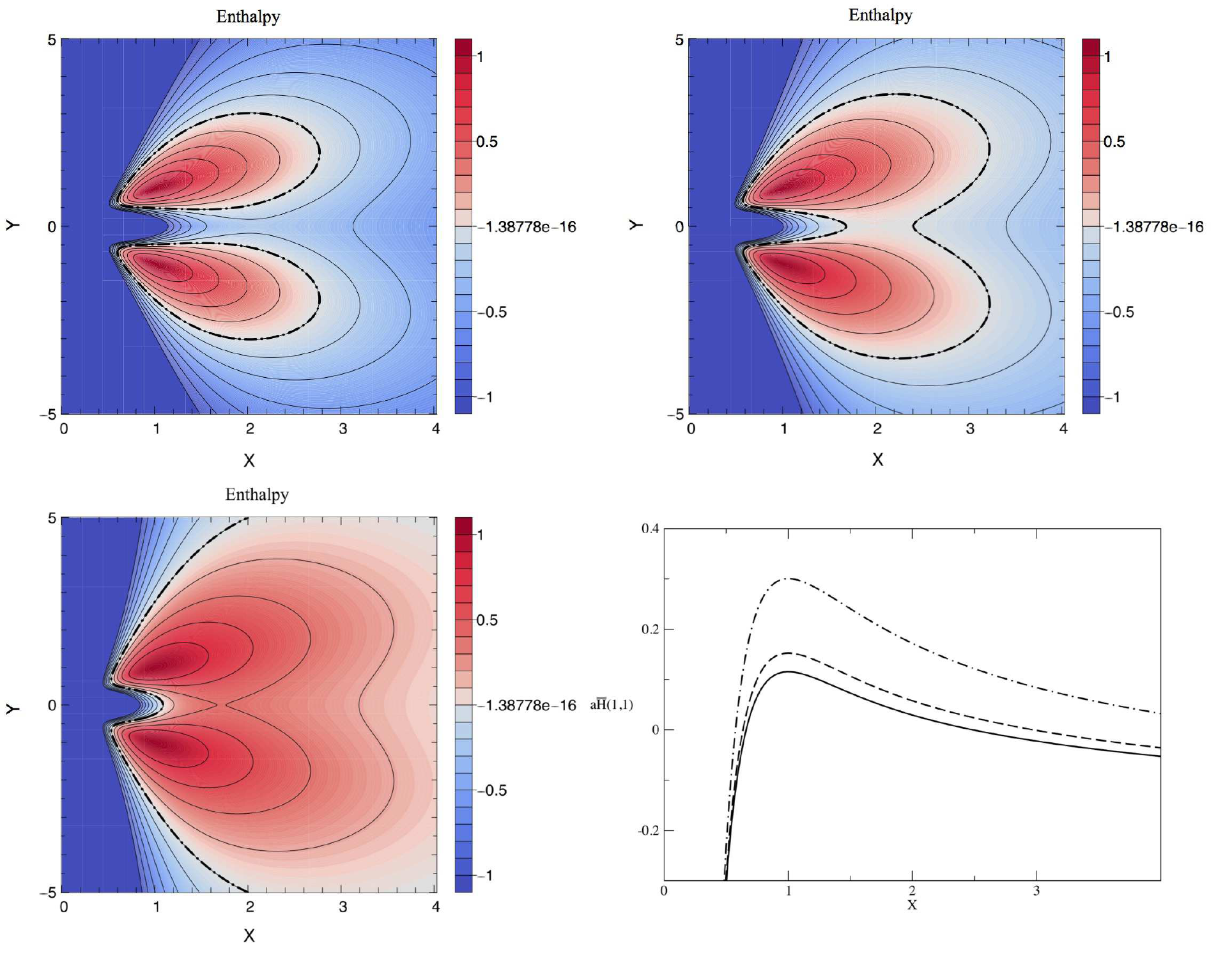}
	\caption{Same as in Figure \ref{f3}, but for off-equatorial tori of Family II and for parameter as given in Table \ref{tab4}. Two differences are included: 1) all the tori are negatively charged, 2) according the value of $\dt$ the morphology of the torus is not the same, the lobes can be linked or not with the equatorial plane.}
  \label{f6}
\end{figure*}

\begin{figure*}
\centering
  \includegraphics[width=1\hsize]{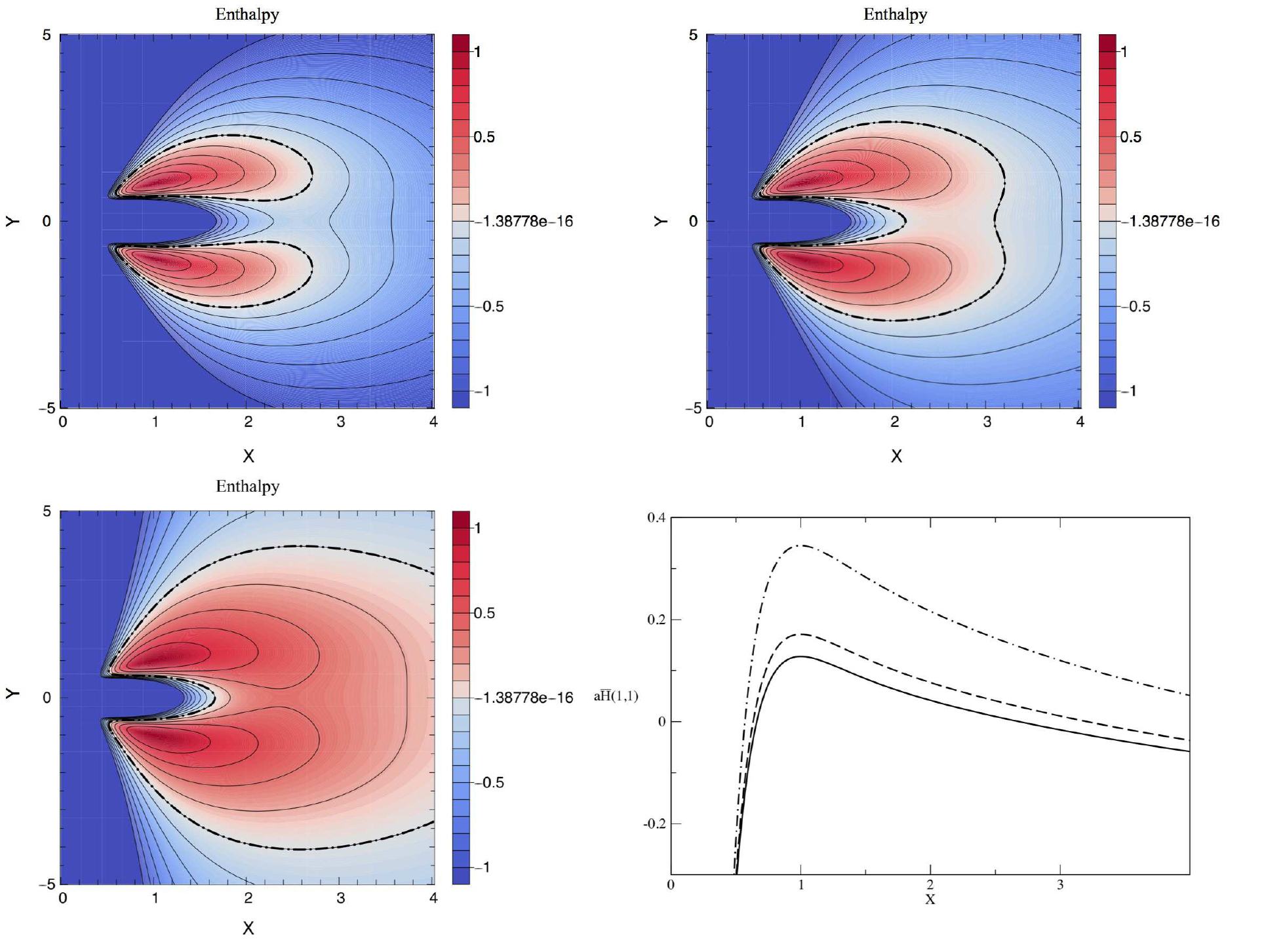}
	\caption{Same as the Figure \ref{f6}, but for Family IV.}
  \label{f7}
\end{figure*}

Now, as above, we produce three maps of enthalpy for three values of $\dt$. We set $r_c=18.75\sqrt{2}/2$, $Y_c=1$, $c=-0.015r_c$ and $K_1=-1$ for family II. The value of $\mu C/\sqrt{GM}$ for each case is given in the left part of Table \ref{tab4}. 
\begin{table}
\begin{center}
\caption{Value of $\mu C/\sqrt{GM}$ for $\dt=0,0.1,0.5$.}
\begin{tabular}{cll}
\hline
\hline
\multicolumn{1}{c}{$\dt$}&\multicolumn{1}{c}{Family II}&\multicolumn{1}{c}{Family IV}\\
\hline
   $0$& $\mu C/\sqrt{GM} \sim -1.633$ &$\mu C/\sqrt{GM} \sim -375$\\
   \hline
   $0.1$ & $\mu C/\sqrt{GM}\sim -1.723$ &$\mu C/\sqrt{GM}\sim -396$\\
   \hline
   $0.5$ & $\mu C/\sqrt{GM}\sim -2.055$ &$\mu C/\sqrt{GM}\sim -471$\\
   \hline
\end{tabular}
\end{center}
\label{tab4}
\end{table}
The results are shown in the Figure \ref{f6}. 
Next, we produced the same figures for family IV with the same parameters except  $c=-0.02r_c$. The value of $\mu C/\sqrt{GM}$ is given in the right part of Table \ref{tab4}. The map linked to this family is shown in Figure \ref{f7}.
We can see, for both families, that for $\dt=0$, there are toroidal off-equatorial structures located above and under the equatorial plane. Given a value of $c$, by increasing $\dt$, the morphology of the solutions changes. The off-equatorial toroidal structures are linked to each other by the equatorial plane. 

\section{Comparison with the Self-Consistent Field method}
\label{sec:SCF}
Whereas the above-described approximation has enabled us to develop a systematic classification across the parameter space of the  constraints for the existence of topologically different toroidal configurations, the adopted limit of an infinitesimally narrow gravitating ring is an idealisation. An astrophysically realistic model will require to take the spatially extended distribution of the torus material and the corresponding pressure and density distribution. In order to relax the mentioned restriction we can compare the analytical description with a corresponding spatially extended configuration constructed numerically. This will allow us to assess the accuracy of the approximation, although the numerical solution does not provide methodology for the classification. To this end we employ the Self-Consistent Field method (SCF) that was developed initially by \cite{OstMa68}.

The SCF approach is based on the integral form of Euler's equation (\ref{eq:bernouilli}). Subsequently, variants of this method were  developed to describe the internal structure of rotating stars, and to explore different figures of equilibrium, namely, the structure of rotating polytropes \citep{Blinnikov75,hachisu86,Eriguchi05}. Originally the method was applied in the non-magnetized case. We have thus programmed a modified version of the SCF method, where we introduce a non-vanishing imposed magnetic term $\cal M$, and we make a comparison with the analytical toy-model approximation.

The numerical set-up is based on an iterative scheme for the density. In our case this translates to solving eq. (\ref{eq:bern_norm}). The iterative loop is initiated by setting the rotation law (e.g., the rigid rotation, or the constant angular momentum density), the radial profile of the specific charge density (e.g., a power-law dependence), the polytropic index ($n=0$ for the incompressible case), the location of  inner and outer edges, and the value of constants $\dt$ and $e$.
 
Additionally, in the magnetized case we introduce the external (dipol-type) magnetic field, and we also set the electric charge density profile within the fluid. Let us note that the enthalpy vanishes at the inner and outer edges of the volume occupied by the rotating fluid. Then, the sequence of steps is as follows:
\begin{itemize}
\item[-] Set an initial profile of density.
\item[-] By employing Poisson's equation, determine the 
corresponding gravitational potential.
\item[-] By using the equation \ref{eq:bern_norm} in the inner and the outer edge, obtain constants $b$ and $c$.
\item[-] Calculate the enthalpy distribution.
\item[-] Assuming the polytropic equation of state for the fluid, the pressure and the enthalpy are linked to the density. Using this relation, the density profile can be updated for the subsequent iteration step.
\item[-] Compare the new density with the initial one. The equilibrium has been achieved if the two profiles coincide with each other within a pre-defined accuracy, otherwise the routine loops back to the first step until the variables reach convergence.
\end{itemize}
For further details on our implementation of the SCF method we refer the reader to a more detailed exposition given elsewhere (Trova et al., 2016, in preparation). Here, for the sake of definiteness we compare the outcome of the above-described procedure with the test $2$ of Tab.\ref{tab2}. We set  $\mu C/\sqrt{GM}=0.010$, $\dt=0.1$, and we assume the same rotation law and the specific charge that are imposed to start the computation. The criterion for achieving the convergence requires that the relative difference between the constant $b$ (or $a$ and $c$) values at subsequent steps reaches $5$ times the machine epsilon precision (around $10^{-6}$ in single precision). We find that the equilibrium is reached, typically, after $10$ to $15$ iterations. We obtained $K_2/\sqrt{GM}=2.45$ and $c/r_c=-0.0247$. These results are consistent with the values used in our analytical approach, see \ref{sec:equa_conf}.

\begin{figure}
\centering
      \includegraphics[width=1\hsize]{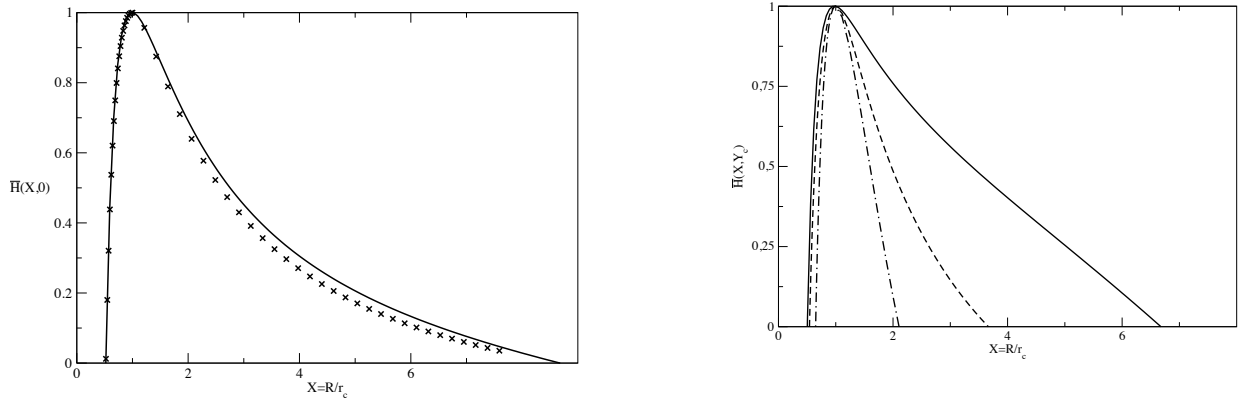}
        \caption{The equatorial radial profile of enthalpy is shown for the two approaches. The result from SCF method in plotted by solid line, and the points from the analytical approach are indicated by crosses.}
      \label{f8}
\end{figure}

The equatorial enthalphy profile is plotted in Figure \ref{f8} for both approaches. We can thus conclude that the equatorial density obtained with our approximation method is close to the corresponding profile reached by the SCF method.

\begin{figure}
\centering
      \includegraphics[width=1\hsize]{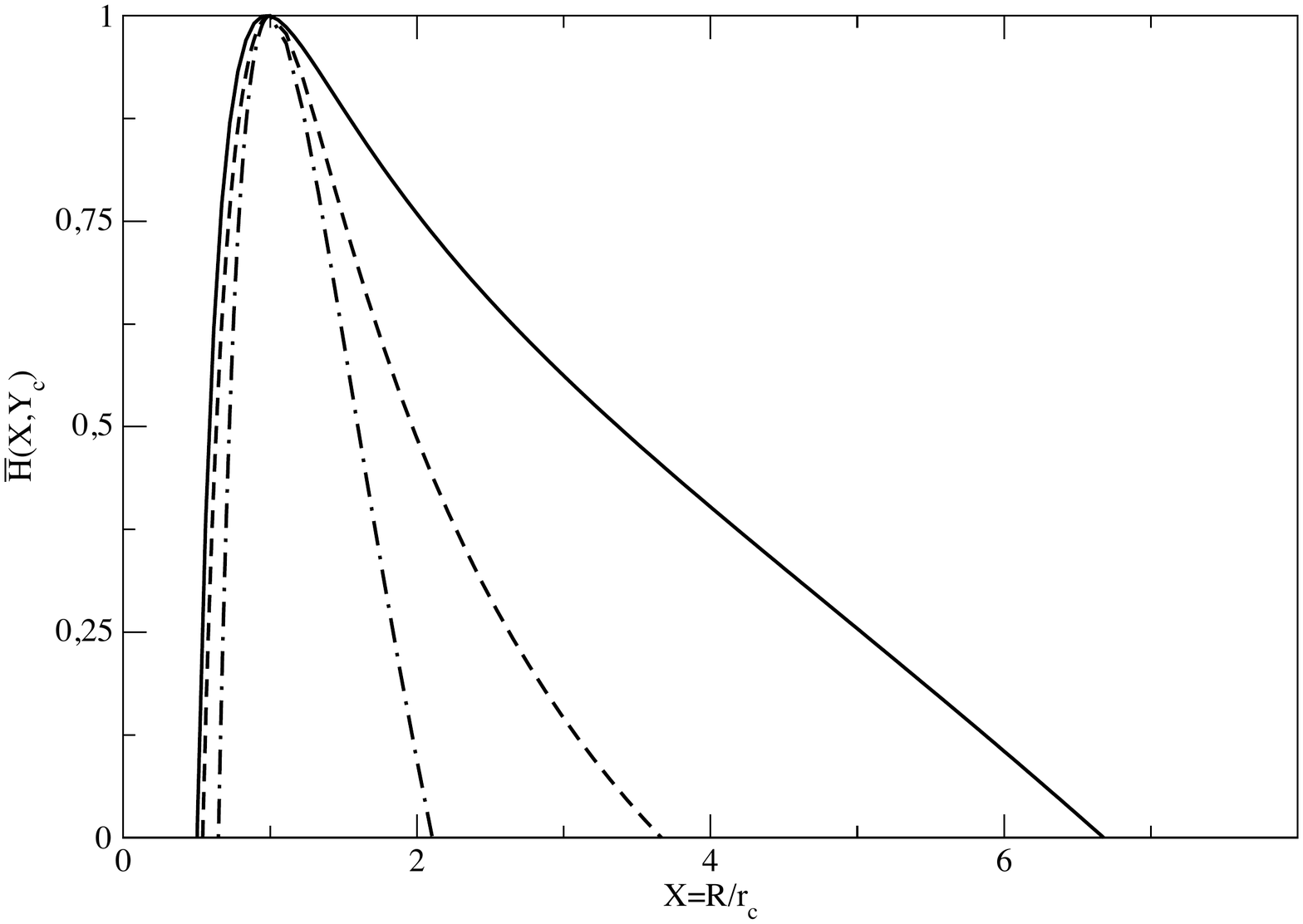}
        \caption{The enthalpy profile is shown analogically to Fig. 8, but for the case of two off-equatorial lobes with $Y_c=Z_c/r_c$. The result from SCF method is plotted by solid line, the one ring approach by dot-dashed line and the double-ring approach is indicated by dashed line.}
      \label{f9}
\end{figure}

We produced the same comparison in the case of the off-equatorial tori. We expect that the results are less consistent with the SCF due to the one ring approximation in the equatorial plane. We performed another test using a double-ring approximation for the gravitational potential. The first ring is located in ($1,Y_c$) and the second one in ($1,-Y_c$).  The enthalpy profile in $Y_C=Z_c/r_c$ (altitude of the maximum of pressure) as a function of $X$ is plotted in Figure \ref{f9}.
As expected, the analytical profile from the one ring approximation comes out quite inaccurate, however, the accuracy is much improved with the double-ring approach. The latter is clearly more precise, although the precision is still not as good as the result from the SCF method. The interesting point is that the morphology does not change. Instead, we find for both cases the two lobes under and above the equatorial plane.

\section{Discussion}
\label{sec:conclusion}
From the conceptual point of view, the presented work introduces a pure topological survey through various globally charged and nonconductive perfect fluid toroidal configurations formed due to complex gravito-electromagnetic interactions. The combined actions of the fluid torus self-gravity together with the electric charge distribution were treated separately in previous papers. Now, being considered together, they represent basic theoretical view on the studied problem. There is, however, a promising astrophysical contextualization, as mentioned in Introduction. In the following, we comment the circumstances essential for the astrophysical on feasibility of the model: (i) global non-zero charge distribution, (ii) ionization as a process for free charges generation, and (iii) estimation of particular physical characteristics of the considered circling matter. In the end, we also present the whole summary of the work.

\subsection{Non-zero charge and ionization}
\label{sec:Nonzero}
The effect of the Lorentz force on the non-vanishing net charge carried by the torus material can help to support the vertical structure of the torus against its own self-gravity, thereby maintaining the geometrical thickness, which would otherwise tend to collapse into the equatorial plane. Non-vanishing net electric charge distribution can develop by various mechanisms depending on the nature of the medium. Free charges are created by ionisation of gas, effect of intense irradiation of dust grains by the central source, and charge exchange within complex plasmas. Charge separation operates in the organised magnetic fields (see \cite{KovarTr14} for further references), so that large scale regions of non-vanishing charge can develop even if the  whole system is globally neutral. On the other hand, the electric charge of central body (Wald's charge) is supposed to be negligibly small because of selective accretion that helps to neutralize the body in the centre.

Active galactic nuclei present an important example of objects where the gas ionisation ranges from small values (at large distance and low luminosity) up to fully ionized medium in the energetic environment near the black hole \citep{Krolik99}. Typically, gas becomes partially ionized when it is subject of irradiation by X-rays, and these become more intense towards the central source. The irradiation mechanism provides free electric charges and ions. The photoionization of the surface of the inner accretion disk (by a hot corona) can be characterized in terms of ionization parameter, $\xi = 4\pi F_X/n_H$, where $F_X$ is the incident X-ray flux and $n_H$ is the hydrogen number density \citep{BallMcDRu11}. A measurement of $\xi$ can therefore provide information on the ionization profile and the density of the environment as functions of illuminating conditions, and thus the number of free charges. Typically, $\xi$ spans from negligible values at the outer edge of the accretion disk up to $\gtrsim10^5$ at a few gravitational radii, where the medium is almost fully ionised \citep{BaFaRo02,RoDuCzCo02}.

However, with respect to electric charge content in realistic conditions, one needs to take also the emergence of dust into account \citep{LaoDrai93,Swany05}. Dust can develop when the grain temperature does not exceed above the sublimation temperature $\sim 1500$ K. Dust grains embedded within the gas will be charged to form complex plasma of a quasar. Charges bind dust to the surrounding partially-ionized gas \citep{IvAkCas16}. The Eddington luminosity ratio is then reduced by a factor equal to dust cross-section (per proton, appropriately weighted for the spectral energy distribution profile) to the Thomson cross section, which comes out of the order of $1:10^3$ \citep{fabian12}.

\subsection{Physical characteristics}

The prime aim of this work is the discussion of a general framework for the combined gravito-electromagnetic actions studied through the enthalpy profiles of the formed tori mapping their geometries. An important challenge for further investigation is the detailed study of other physical characteristics (such as pressure, mass density, temperature, specific charge, etc.) describing more of the tori microphysics. This study requires precise adjustment of thermodynamical relations, representing a delicate problem. It means to choose reliable pressure-mass density and pressure-temperature relations. Moreover, the desired profiles of physical characteristics are sensitive to the central mass, the magnetic field strength, and of the torus size, etc. A detailed quantitative discussion is beyond the scope of the present paper (a separate work in progress).

In order to give a rough view of the physical characteristics throughout the tori, we refer to our recent work \citep{KovarTr16}, where the same problem of electrically charged and nonconductive perfect fluid toroidal configurations in central gravitational and magnetic dipolar fields is discussed within general relativity framework; there, however, without the self-gravity of the tori. In that paper, we concluded that the circling fluid with constant specific angular momentum, under polytropic equation of state and with pressure being related to the temperature by the ideal gas relation form astrophysically relevant equatorial and off-equatorial tori with feasible pressure, mass density and temperature profiles; in the presented sample cases reaching their maxima in centres in intervals $P_{\rm max}\simeq10^{13}$--$10^{16}$ Pa, $\rho_{\rm max}\simeq 10^2-10^4\,{\rm kg.m^{-3}}$, $T_{\rm max}\simeq10^7-10^8\,{\rm K}$, and with average specific charges $10^{-11}$ of the proton one. We also showed that such tori must be relatively tiny in comparison with the central object. Being located close to the central object at radii $\sim10GM/c^2$, their cross-sectional size is very small (a slender torus approximation). However, let us note that specific numerical values of the physical quantities can be tuned over a wide span in our general scheme because the initial assumptions contain a number of degrees of freedom for which we lack clear observational constraints.

\section{Conclusion}
In this paper, we discussed the impact of the self-gravity on the conditions of existence of charged fluid tori and their morphology. The fluid, whose particles carry electrical charges, was assumed to be perfect and incompressible, the latter due to easier handling with equations. It was influenced by its own gravitational field, and by the gravitational potential and the dipolar magnetic field of the central mass. We base our study on the work of \cite{slany13}. We proceeded in the same way but included self-gravity. We analysed the Euler's equation to find stationary toroidal configurations for two families of specific charge.

The first interesting result is that the condition of existence of the tori changes with the strength of self-gravity, as characterized by the parameter $\dt$. The parameters allowing the existence of equatorial tori change with the value of $\dt$. For off-equatorial tori, as in the case without self-gravity, there is always a solution which is negatively charged. For both these families, for $K_2>0$ (positive direction of motion) and for $\mu>0$ (orientation of the magnetic field), positively charged tori do not exist. The off-equatorial tori would have positive charge only for $K_2<0$ or $\mu<0$.

Another interesting result is the impact of self-gravity on the charge of the equilibrium torus. As we saw in section \ref{sec:equa_conf}, the sign of the total charge can change. On the other hand, the morphology of tori is similar to the non-self-gravitating case. We found the toroidal configuration, the closed isobars with cusps, and the toroidal off-equatorial configurations. The maximum of pressure, however, rises with the value of $\dt$ and the torus becomes thicker, which makes sense because higher gravity implies higher pressure to balance the gravitational and electric forces. Finally, the last interesting point is the possibility for these two families to have solutions in rigid rotation, which exist for a self-gravitating torus without a spherical gravitational and a dipolar magnetic field too \citep{hachisu86}, but not for non self-gravitating tori with the specific charge distributions described in \cite{slany13}, from which we analysed two exemplary of four possibilities in the complete classification. 

The above described approach in this paper can serve as a useful test bed for comparisons with other methods. In particular, it provides us with a better insight into conditions that define the form of the electrically charged configurations. The method allows us to produce a relatively precise approximation to their structure, taking self-gravity 
of the fluid into account. While the precision of the method can be verified numerically in the selected cases, e.g. by employing the SCF scheme, the closed analytical form provides a way to set constraints on the existence of different configurations.

While the combination of a large-scale organized (dipole-like) magnetic component and a non-vanishing net charge of the fluid are required to allow the emergence and stability of toroidal structures outside the equatorial plane, self-gravity acts against them. It was therefore interesting to verify, as we did in this paper, that the resulting lobes of matter above and below the equatorial plane can persist even when self-gravitational force is taken into account.

\section*{Acknowledgments}
It is a pleasure to thank Prof. J-M Hur\'e for his help and valuable advice. The authors would also like to thank the referee for useful comments and suggestions. AT is grateful to Relativistic Astrophysics Group and the Academy of Sciences in Prague. We thank the Czech Science Foundation Center of Excellence titled ``Albert Einstein Center for Gravitation and Astrophysics'' (No. 14-37086G) and the COST Action MP1304 on Exploring fundamental physics with compact stars (No. LD15061).

\bibliographystyle{mn2e}
\bibliography{Toy_model}

\appendix
\section{A: Differential equations for the magnetic potential}
\label{appendixA}
In cylindrical coordinates ($R$,$\phi$,$Z$) we have, 
\begin{equation}
\vec{B}=\mathbf{rot} \vec{A} = \left|
      \begin{aligned}
        -\frac{\partial \aphi }{\partial Z}\\
        0 \\
        \frac{1}{R} \frac{\partial (R\aphi)}{\partial R}\\
      \end{aligned}
    \right..
\end{equation}
So the Lorentz force is given by
\begin{equation}
\frac{\vec{\cal{L}}}{\pf}=\frac{\pe \vec{v} \wedge \vec{B}}{\pf} =q v_{\phi}\vec{\ephi} \wedge \vec{B}= \left|
      \begin{aligned}
        \frac{qv_{\phi}}{R} \frac{\partial (R\aphi)}{\partial R}\\
        0 \\
        qv_{\phi}\frac{\partial \aphi }{\partial Z}\\
      \end{aligned}
    \right..
\end{equation}
We need $\vec{\nabla}{\cal{M}}=-\vec{\cal{L}}/\pf$. Then
\begin{equation}
-\left|
      \begin{aligned}
        \frac{qv_{\phi}}{R} \frac{\partial (R\aphi)}{\partial R}\\
        0 \\
        \frac{qv_{\phi}}{R}\frac{\partial (R\aphi) }{\partial Z}\\
      \end{aligned}
    \right. \quad = \quad 
    \left|
      \begin{aligned}
        \frac{\partial \cal{M}}{\partial R}\\
        0 \\
        \frac{\partial \cal{M} }{\partial Z}\\
      \end{aligned}
    \right..
\label{eq:condition}
\end{equation}

\section{B: Value of the first and second derivatives of the gravitational potential of the ring in $(1,Y_c)$}
\label{appendixB}
The normalized gravitational potential is given by 
\begin{equation}
\label{eq:normk}
\tpsig = -\frac{1}{\pi} \sqrt{\frac{1}{X}}k\elik(k) \quad \text{with} \quad k=\frac{2\sqrt{X}}{\sqrt{(1+X)^2+Y^2+\lambda^2}}.
\end{equation}
The first derivative of $\tpsig$ with respect to the normalized radius $X$ can be written as
\begin{equation}
\label{eq:derivative1R}
\frac{\partial \tpsig}{\partial X}=-\frac{1}{\pi}\left( \sqrt{\frac{1}{X}}\frac{\partial k}{\partial X}\frac{\elie(k)}{1-k^2} - \frac{1}{2} \frac{\sqrt{X}}{X^2} k\elik(k) \right) \quad \text{where} \quad \frac{\partial k}{\partial X}=\frac{k}{2X}-k^3 \left(\frac{1+X}{4X}\right),
\end{equation}
with $k$ being given by the equation (\ref{eq:normk}) and $\elie(k)$ the complete integral elliptic of the second kind \citep{gradryz65}.
In the maximum of pressure $(1,Y_c)$, the equation (\ref{eq:derivative1R}) becomes
\begin{equation}
\frac{\partial \tpsig}{\partial X}=-\frac{k_*}{2\pi} \left[\elie(k_*)-\elik(k_*)\right] \quad \text{where} \quad k_*=\frac{2}{\sqrt{4+\lambda^2+Y_c^2}}.
\end{equation} 
The second derivative with respect to $X$ is given by 
\begin{equation}
\label{eq:derivative2R}
\pi \frac{\partial^2 \tpsig}{\partial X^2}=\elik(k)\left[\sqrt{\frac{1}{X}}\frac{1}{kk'^2}\left(\frac{\partial k}{\partial X}\right)^2-\frac{3k\sqrt{X}}{4X^3}\right]
-\frac{\elie(k)}{k'^2}\left\{-\frac{\sqrt{X}}{X^2}\frac{\partial k}{\partial X}+\sqrt{\frac{1}{X}} \left[\frac{\partial^2 k}{\partial X^2}+\frac{1+k^2}{kk'^2}\left(\frac{\partial k}{\partial X}\right)^2 \right] \right\},
\end{equation}
with $k'^2=1-k^2$ and
\begin{equation}
\frac{\partial^2 k}{\partial X^2}=\frac{1}{2X}\frac{\partial k}{\partial X}-\frac{k}{2X^2}-3k^2\frac{\partial k}{\partial X}\left(\frac{1+X}{4X}\right)+\frac{k^3}{4X^2}.
\end{equation}
In $(1,Y_c)$, the equation (\ref{eq:derivative2R}) becomes
\begin{equation}
\frac{\partial^2 \tpsig}{\partial X^2}=-\frac{k_*}{4\pi}\left[\left(\frac{k_*^2-2}{{k'}_*^2}-2k_*^2\right)\elie(k_*)+(k_*^2+2)\elik(k_*)\right],
\end{equation}
with ${k'_*}^2=1-{k_*}^2$. The first derivative with respect to $Y$ is
\begin{equation}
\label{eq:derivative1Z}
\pi \frac{\partial \tpsig}{\partial Y}=\sqrt{\frac{1}{X}}\frac{Yk^3}{4X}\frac{\elie(k)}{k'^2}.
\end{equation}
In $(1,Y_c)$, the equation (\ref{eq:derivative1Z}) becomes
\begin{equation}
\pi \frac{\partial \tpsig}{\partial Y}=\frac{Y_ck_*^3}{4}\frac{\elie(k_*)}{{k'_*}^2}.
\end{equation}
The second derivative with respect to $Y$ is written as
\begin{equation}
\label{eq:derivative2Z}
-\pi \frac{\partial^2 \tpsig}{\partial Y^2}=\sqrt{\frac{1}{X}}\frac{\partial^2 k}{\partial X^2}\frac{\elie(k)}{k'^2}
+\left(\frac{\partial k}{\partial Y}\right)^2\frac{1}{kk'^2}\left(\elie(k)\frac{1+k^2}{k'^2}-\elik(k)\right)\sqrt{\frac{1}{X}},
\end{equation}
with 
\begin{equation}
\frac{\partial k}{\partial Y}=-\frac{Yk^3}{4X},
\end{equation}
and
\begin{equation}
\frac{\partial^2 k}{\partial Y^2}=-\frac{k^3}{4X}-\frac{3Yk^2}{4X}\frac{\partial k}{\partial Y}.
\end{equation}
In $(1,Y_c)$, the equation (\ref{eq:derivative2Z}) becomes
\begin{equation}
-\pi \frac{\partial^2 \tpsig}{\partial Y^2}=\frac{\elie(k_*)}{{k'}_*^2}\frac{{k}_*^3}{4}\left[-1+\frac{Y_c^2{k}_*^2}{4}\left(\frac{4-2{k}_*^2}{1-{k}_*^2}\right)\right]-\frac{\elik(k_*){k}_*^5Y_c^2}{16{k'}_*^2},
\end{equation}
and in $Y_c=0$
\begin{equation}
\frac{\partial^2 \tpsig}{\partial Y^2}=\frac{{k_*}^3}{4\pi}\frac{\elie(k_*)}{{k'_*}^2}.
\end{equation}
To finish, we calculate the second-order mixed derivatives of $\tpsig$. We differentiate the equation (\ref{eq:derivative1Z}) with respect to $X$.
\begin{equation}
\label{eq:derivativeRZ}
\pi \frac{\partial^2 \tpsig}{\partial X \partial Y}=\sqrt{\frac{1}{X}}\frac{k^2Y}{4X}\left\{\frac{\elie(k)}{{k'}^2}\left[-\frac{3k}{2R}+\frac{\partial k}{\partial X}\frac{(4-2k^2)}{k'^2}\right]\right.
-\left.\frac{\partial k}{\partial X}\frac{\elik(k)}{{k'}^2}\right\} \nonumber.
\end{equation}
In $(1,Y_c)$, the equation (\ref{eq:derivativeRZ}) writes 
\begin{equation}
\pi \frac{\partial^2 \tpsig}{\partial X \partial Y}=\frac{k_*^3Y_c}{8}\left\{\frac{\elie(k_*)}{{k'_*}^2}[1-2k^2]-\elik(k_*)\right\}.
\end{equation}

\section{C: Expression of the derivatives of the enthalpy.}
\label{appendixC}
\begin{itemize}
\item Family II\\
According to the equation (\ref{eq:bern_norm}), we have
\begin{equation}
a\frac{\partial \th}{\partial X}= -\frac{X}{\left(X^2+Y^2\right)^\frac{3}{2}}-\dt \frac{\partial \tpsig}{\partial X}+bX^{2K_1-1}+eX^{-2K_1}\left(X^2+Y^2\right)^{3K_1/2-7/4}(2Y^2-X^2),
\end{equation}
\begin{equation}
a\frac{\partial \th}{\partial Y}= -\frac{Y}{\left(X^2+Y^2\right)^\frac{3}{2}}-\dt \frac{\partial \tpsig}{\partial Y}-3eX^{-2K_1+1}Y\left(X^2+Y^2\right)^{3K_1/2-7/4},
\end{equation},
\begin{flalign}
a\frac{\partial^2 \th}{\partial X^2}= &\frac{2X^2-Y^2}{\left(X^2+Y^2\right)^\frac{5}{2}}-\dt \frac{\partial^2 \tpsig}{\partial X^2}+(2K_1-1)bX^{2K_1-2}\\ \nonumber
&-\frac{e}{2} \frac{\left(X^2+Y^2\right)^{3K_1/2-11/4}}{X^{2K_1+1}}(-8K_1X^2Y^2+8K_1Y^4+2K_1X^4-3X^4+18X^2Y^2),
\end{flalign}
\begin{equation}
a\frac{\partial^2 \th}{\partial Y^2}= -\frac{X^2-2Y^2}{\left(X^2+Y^2\right)^\frac{5}{2}}-\dt \frac{\partial^2 \tpsig}{\partial Y^2}-\frac{3e}{2} \frac{\left(X^2+Y^2\right)^{3K_1/2-11/4}}{X^{2K_1-1}}(2X^2-5Y^2+6K_1Y^2),
\end{equation}
and
\begin{equation}
a\frac{\partial^2 \th}{\partial X\partial Y}= \frac{3XY}{\left(X^2+Y^2\right)^\frac{5}{2}}-\dt \frac{\partial^2 \tpsig}{\partial X \partial Y}-\frac{3e}{2} \frac{\left(X^2+Y^2\right)^{3K_1/2-11/4}}{X^{2K_1}}Y(2K_1X^2-4K_1Y^2-5X^2+2Y^2).
\end{equation}

\item Family IV\\
\begin{equation}
a\frac{\partial \th}{\partial X}= -\frac{X}{\left(X^2+Y^2\right)^\frac{3}{2}}-\dt \frac{\partial \tpsig}{\partial X}+bX^{2K_1-1}+e\left(\frac{X}{\sqrt{X^2+Y^2}}\right)^{3(1-K_1)}\frac{2Y^2-X^2}{\left(X^2+Y^2\right)^\frac{5}{2}}X^{K_1},
\end{equation}
\begin{equation}
a\frac{\partial \th}{\partial Y}= -\frac{Y}{\left(X^2+Y^2\right)^\frac{3}{2}}-\dt \frac{\partial \tpsig}{\partial Y}-3e\left(\frac{X}{\sqrt{X^2+Y^2}}\right)^{3(1-K_1)}\frac{X^{K_1+1}Y}{\left(X^2+Y^2\right)^\frac{5}{2}},
\end{equation},
\begin{flalign}
a\frac{\partial^2 \th}{\partial X^2}= &\frac{2X^2-Y^2}{\left(X^2+Y^2\right)^\frac{5}{2}}-\dt \frac{\partial^2 \tpsig}{\partial X^2}+(2K_1-1)bX^{2K_1-2}\\ \nonumber
&-e \frac{\left(X^2+Y^2\right)^{3K_1/2-5}}{X^{2K_1-2}}(15X^2Y^2-6Y^4-3X^4-4K_1X^2Y^2+4K_1Y^4-K_1X^4),
\end{flalign}
\begin{equation}
a\frac{\partial^2 \th}{\partial Y^2}= -\frac{X^2-2Y^2}{\left(X^2+Y^2\right)^\frac{5}{2}}-\dt \frac{\partial^2 \tpsig}{\partial Y^2}-3e \frac{\left(X^2+Y^2\right)^{3K_1/2-5}}{X^{-2K_1-4}}(-7Y^2-X^2+3K_1Y^2),
\end{equation}
and
\begin{equation}
a\frac{\partial^2 \th}{\partial X\partial Y}= \frac{3XY}{\left(X^2+Y^2\right)^\frac{5}{2}}-\dt \frac{\partial^2 \tpsig}{\partial X \partial Y}-3e \frac{\left(X^2+Y^2\right)^{3K_1/2-5}}{X^{-3-2K_1}}Y(5K_1X^2+2K_1Y^2-4X^2+4Y^2).
\end{equation}
\end{itemize}

\section{D: Description of the $F_1$, $F_2$, $G_1$ and $G_2$.}
\label{appendixD}
In this appendix, we give the explicit form of functions $F_1$, $F_2$, $G_1$ and $G_2$ which appear in section \ref{sec:off_cond} for both distributions. In each function, the first, second and mixed derivative of $\psig$ are taken in $X=1$ and $Y=Y_c$.
\begin{itemize}
\item Family II
\begin{equation}
F_1=\frac{[(10K_1+5)+Y_c^4(16K_1-4)+Y_c^2(8K_1+4)]}{4(1^2+Y_c^2)^2},
\end{equation}
\begin{flalign}
F_2=&-\frac{3}{2}\frac{(2-Y_c^2)}{(1+Y_c^2)^2}\left[\frac{\partial \tpsig}{\partial X}-\frac{1-2Y_c^2}{3Y_c}\frac{\partial \tpsig}{\partial Y}\right]-\frac{\partial^2 \tpsig}{\partial X^2}\\ \nonumber
&-\left[\frac{\partial \tpsig}{\partial X}-\frac{1}{Y_c}\frac{\partial \tpsig}{\partial Y}\right]\frac{Y_c^2(18-8K_1)+8K_1Y_c^4+(2K_1-3)}{2(1+Y_c^2)^2},
\end{flalign}
\begin{equation}
G_1=\frac{3}{2}\frac{{4K_1-2}Y_c^2}{(1+Y_c^2)^2}(2K_1+1)(3K_1-1).
\end{equation}
The function $G_2$ is a combination of various functions which depend, as $F_2$, on $Y_c$, the first, second and mixed derivatives of the self-gravitational potential.
\begin{equation}
\label{eq:G2}
G_2=b(F_1 H_1+F_2 H_2-2 H_3 H_4)+\dt(F_2 H_1-H_4^2),
\end{equation}
where
\begin{equation}
H_1=\frac{3}{2}\frac{(1-2Y_c^2)}{(1+Y_c^2)^2}\left[\frac{\partial \tpsig}{\partial X}-\frac{1-2Y_c^2}{3Y_c}\frac{\partial \tpsig}{\partial Y}\right]-\frac{\partial^2 \tpsig}{\partial Y^2}
-\frac{3(6K_1Y_c^2-5Y_c^2+2)}{2(1+Y_c^2)^2}\left[\frac{\partial \tpsig}{\partial X}-\frac{1}{Y_c}\frac{\partial \tpsig}{\partial Y}\right],
\end{equation}
\begin{equation}
H_2=\frac{3Y_c^2}{4(1+Y_c^2)^2}(6K_1-1),
\end{equation}
\begin{equation}
H_3=\frac{3Y_c}{4(1+Y_c^2)^2}[(1+2K_1)+Y_c^2(2-4K_1)],
\end{equation}
and 
\begin{equation}
H_4=-\frac{9Y_c}{2(1+Y_c^2)^2}\left[\frac{\partial \tpsig}{\partial X}-\frac{1-2Y_c^2}{3Y_c}\frac{\partial \tpsig}{\partial Y}\right]-\frac{\partial \tpsig}{\partial X \partial Y}
-\frac{3Y_c(-4K_1Y_c^2+2K_1+2Y_c^2-5)}{2(1+Y_c^2)^2}\left[\frac{\partial \tpsig}{\partial X}-\frac{1}{Y_c}\frac{\partial \tpsig}{\partial Y}\right].
\end{equation}

\item Family IV\\
The functions are defined as follows,
\begin{equation}
F_1=\frac{[(5K_1+1)+Y_c^4(8K_1-8)+Y_c^2(4K_1+8)]}{(1+Y_c^2)^2},
\end{equation}
\begin{flalign}
F_2=&-\frac{3}{2}\frac{(2-Y_c^2)}{(1+Y_c^2)^2}\left[\frac{\partial \tpsig}{\partial X}-\frac{1-2Y_c^2}{3Y_c}\frac{\partial \tpsig}{\partial Y}\right]-\frac{\partial^2 \tpsig}{\partial X^2}\\ \nonumber
&-\left[\frac{\partial \tpsig}{\partial X}-\frac{1}{Y_c}\frac{\partial \tpsig}{\partial Y}\right]\frac{Y_c^2(15-4K_1)-Y_c^4(6-4K_1)-(3-K_1)}{2(1+Y_c^2)^2},
\end{flalign}
\begin{equation}
G_1=\frac{3Y_c^2}{(1+Y_c^2)^2}(-2-4K_1+3K_1^2).
\end{equation}
The function $G_2$ is given by the equation (\ref{eq:G2}) with the functions $H_1$, $H_2$, $H_3$ and $H_4$ defined as follows.
\begin{equation}
H_1=\frac{3}{2}\frac{(1-2Y_c^2)}{(1+Y_c^2)^2}\left[\frac{\partial \tpsig}{\partial X}-\frac{1-2Y_c^2}{3Y_c}\frac{\partial \tpsig}{\partial Y}\right]-\frac{\partial^2 \tpsig}{\partial Y^2}
-\frac{3(1-7Y_c^2+3K_1Y_c^2)}{2(1+Y_c^2)^2}\left[\frac{\partial \tpsig}{\partial X}-\frac{1}{Y_c}\frac{\partial \tpsig}{\partial Y}\right],
\end{equation}
\begin{equation}
H_2=\frac{3Y_c^2}{2(1+Y_c^2)^2}(3K_1-5),
\end{equation}
\begin{equation}
H_3=\frac{3Y_c}{2(1+Y_c^2)^2}[(K_1-1)+Y_c^2(4-2K_1)],
\end{equation}
and 
\begin{equation}
H_4=-\frac{9Y_c}{2(1+Y_c^2)^2}\left[\frac{\partial \tpsig}{\partial X}-\frac{1-2Y_c^2}{3Y_c}\frac{\partial \tpsig}{\partial Y}\right]-\frac{\partial \tpsig}{\partial X \partial Y}
-\frac{3Y_c(-2K_1Y_c^2+K_1+4Y_c^2-4)}{2(1+Y_c^2)^2}\left[\frac{\partial \tpsig}{\partial X}-\frac{1}{Y_c}\frac{\partial \tpsig}{\partial Y}\right].
\end{equation}
\end{itemize}

\end{document}